\documentclass[aps,prl,twocolumn,fleqn,showpacs,superscriptaddress,longbibliography]{revtex4-2}
\usepackage{mathtools,amssymb,graphicx,units}

\usepackage{color}
\definecolor{darkblue}{RGB}{0,0,127}
\usepackage[plainpages=false,pdfpagelabels,colorlinks=true,linkcolor=darkblue,urlcolor=darkblue,citecolor=darkblue,%
            pdftitle={Title},pdfauthor={},pdfdisplaydoctitle=true,pdfduplex=DuplexFlipLongEdge]{hyperref}

\usepackage[charter,expert]{mathdesign}
\usepackage[scaled=0.96]{XCharter}
\begin{document}
\title{Sensing coherent phonon dynamics in solids with delayed even harmonics}
\author{Jinbin Li}
\email{jinbin_phy@nxu.edu.cn}
\affiliation{Max-Planck-Institut f{\"u}r Physik komplexer Systeme, N{\"o}thnitzer Stra{\ss}e 38, 01187 Dresden, Germany}
\affiliation{School of Nuclear Science and Technology, Lanzhou University, Lanzhou 730000, China}
\affiliation{School of Physics, Ningxia University, Yinchuan, 750021, China}
\author{Ulf Saalmann}
\email{us@pks.mpg.de}
\affiliation{Max-Planck-Institut f{\"u}r Physik komplexer Systeme, N{\"o}thnitzer Stra{\ss}e 38, 01187 Dresden, Germany}
\author{Hongchuan Du}
\email{duhch@lzu.edu.cn}
\affiliation{School of Nuclear Science and Technology, Lanzhou University, Lanzhou 730000, China}
\author{Jan Michael Rost}
\email{rost@pks.mpg.de}
\affiliation{Max-Planck-Institut f{\"u}r Physik komplexer Systeme, N{\"o}thnitzer Stra{\ss}e 38, 01187 Dresden, Germany}

\date{\today}
\begin{abstract}\noindent
High harmonics have emerged as a powerful ultrafast probe of phonon dynamics and electron–phonon interactions in solids, with most studies focusing on odd harmonics.
 Here, in a pump–probe setup with variable delay, we theoretically investigate how even harmonics reveal coherent phonon dynamics. 
 If pump and probe pulses overlap temporally, the spatial interference effect resulting from a non-coaxial pump-probe setup suppresses harmonic yields. 
 At longer delays, odd-harmonic yields oscillate in phase at the optical phonon frequency, whereas even harmonics exhibit order-dependent phase-shifted oscillations. We identify a responsive range of even harmonic orders, in which the delay of yield oscillations is highly sensitive to subtle features of phonon dynamics and electron–electron interactions. 
 Our findings highlight the potential of even harmonics to elucidate microscopic effects in systems with dynamically broken inversion symmetry.
\end{abstract}

\maketitle

\textit{Introduction.} Phonons, quantized lattice vibrations, play a fundamental role in shaping various material properties. Serving as primary mediators, they govern the thermal and acoustic transmission in solids~\cite{Böer2018,SoundHeatRevolutions,Phononengineered,QuantumHeat2021}. Phonons influence electronic properties via electron-phonon interactions that significantly affect charge transport in metals and semiconductors~\cite{Neophytou2020,leeElectronPhononPhysics2023,PhysRevB.109.045149}. They also contribute to the formation of Cooper pairs crucial for superconductivity at low temperatures~\cite{RevModPhys.71.S313,RevModPhys.96.025002,KAUR20213612}. Moreover, phonons are vital to optical properties, evident in phenomena such as Raman scattering~\cite{RamanCrystals} and phonon-assisted absorption~\cite{PhysRevB.82.085303,PhysRevLett.108.167402,Patrick_2014}. A deeper understanding of phonon dynamics and electron-phonon interactions is crucial for the development of materials with targeted properties and improvements in electronic device efficiency.

In the last 15 years, the study of high-order harmonic generation (HHG) from solids has made significant advances~\cite{review1,review2,review3,review4,review5,review6}. Solid HHG-based techniques for probing band structures~\cite{PhysRevLett.115.193603,PhysRevA.104.063109,laninMappingElectronBand2017,Wang_2016,math10224268}, Berry curvatures~\cite{luu2018measurement}, valence electron potentials~\cite{lakhotia2020laser}, intraband and interband dynamics~\cite{Linking_g_s,PhysRevA.109.063511} 
as well as simple models for studying basic processes \cite{wugh+15,TDDFT_1,naci+19,li_SBEs_2019,yusa+22} 
have been developed. Recent progress has also highlighted the potential of high-order harmonics to trace coherent phonon dynamics. Harmonic-yield~\cite{refId0,ex_ph_VO2} and energy~\cite{zhangHighharmonicSpectroscopyProbes2024} oscillations at phonon frequencies have been observed in pump-probe experiments. In experimental and subsequent theoretical investigations~\cite{refId0,th_ph_A,zhangHighharmonicSpectroscopyProbes2024,PhysRevLett.133.156901}, the yield and energy oscillations of harmonics have been attributed to variations in the electronic structure induced by phonons. With this knowledge, HHG-based methods were introduced to investigate phonon dynamics~\cite{zhangHighharmonicSpectroscopyProbes2024} and reconstruct electron-phonon coupling matrices~\cite{PhysRevLett.133.156901}, complemented by first-principles calculations and X-ray diffraction, respectively.
However, most studies have focused on odd harmonics, thereby missing out
on the potential of even harmonics to probe subtle dynamical features. 

A few pioneering studies on reconstructing band structures~\cite{PhysRevLett.115.193603} and probing intraband and interband dynamics~\cite{Linking_g_s} have demonstrated the diagnostic power of even harmonics. This sensitivity could provide a route to systematically probe phonon-induced perturbations, such as the impact of the probe-pulse on the phonon dynamics, which has not been addressed to date.

To render this route explicit, we formulate a minimal model that parameterizes the essential features of pump-probe experiments and is suitable for revealing the generic features that even harmonics exhibit upon phonon-induced perturbations. We validate our model by qualitatively reproducing oscillations of harmonics observed in pump-probe experiments, considering both microscopic responses and spatial interference induced by non-coaxial pump-probe setups. In the next step, we analyze the impact of this interference on harmonics. Subsequently, we determine the distinct behavior of odd and even harmonics in pump-probe measurements and elucidate the underlying nature of these differences. Finally, we reveal that even harmonics are particularly sensitive to the impact of the probe pulse on phonon dynamics and electron-electron interaction.

\begin{figure}[tbp]
\includegraphics[width=8.5cm]{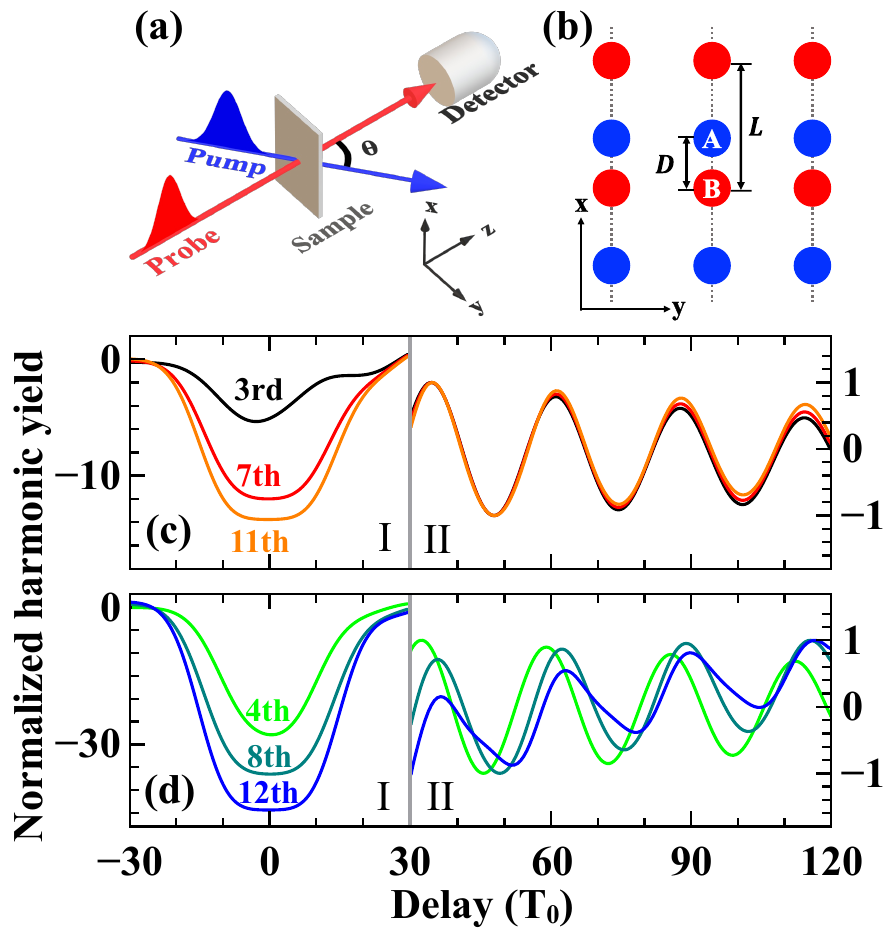} 
 \caption{(a): Sketch of the non‑coaxial pump–probe configuration. The angle $\theta$ between the propagation directions of the pump and probe pulses is indicated. 
 (b): Diagram of the model solid consisting of parallel one‑dimensional diatomic chains. The two different atoms are labeled A and B, with lattice constant $L$ and inter‑atomic distance $D$. (c and d): Integrated harmonic yields as a function of pump–probe delay for selected odd and even harmonics obtained from the TCD calculations. A negative delay corresponds to the pump pulse arriving before the probe pulse. Regions labeled “I” and “II” denote temporal overlap and temporal separation between the pump and probe pulses, respectively. Note that the scale for curves in I and II are very different. For better comparison, the curves are vertically offset and normalized such that their extrema in region II are $+1$ and $-1$, respectively.
}
\label{fig:yield}
\end{figure}

\textit{Methods.} We simulate the pump-probe measurement illustrated in Fig.\,\ref{fig:yield}a, employing a non-coaxial configuration similar to experiments~\cite{refId0,ex_ph_VO2,zhangHighharmonicSpectroscopyProbes2024}. The sample lies in the $x$-$y$-plane, with both pump and probe pulses polarized along the $x$-axis. The probe pulse propagates along the $z$-direction, while the pump pulse travels along an axis inclined by an angle $\theta$ relative to the $z$-axis. The solid sample is modeled as an array of parallel one-dimensional diatomic chains without inter-chain coupling, as shown in Fig.\,\ref{fig:yield}b. 

The laser–matter interaction within a single atomic chain is treated in dipole approximation using time-dependent density functional theory (TDDFT) with classical nuclear motion, for details see part 1 of EndMatter. To align with experimental measurements, phenomenological damping terms are included in both the electronic and lattice equations of motion. 
The parameters of the chain are chosen so that the ground state does not possess inversion symmetry, thereby permitting the generation of even harmonics. Each unit cell contains two nuclei (A and B): one with an effective charge of $Q_\mathrm{A}{=}{+}3$ and a mass of $M_\mathrm{A}{=}16{\times}1836$\,a.u.\ and the other with an effective charge of $Q_\mathrm{B}{=}{+}1$ and a mass of $M_\mathrm{B}{=}20{\times}1836$\,a.u., respectively. Including the electrons, the system is neutral in both charge and spin. The lattice constant is set to $L\,{=}\,6.4$\,a.u., and the equilibrium separation $D_\mathrm{eq}$ between A and B is obtained self-consistently, see EndMatter part 1. The chain contains an acoustic and an optical phonon branch, while only phonon modes at the $\Gamma$ point are excited under the dipole approximation. The acoustic $\Gamma$ mode corresponds to rigid lattice translations, rendering the optical $\Gamma$ mode dominant in modulating the electronic structure and dynamics.

Due to the differing propagation directions of the pump and probe pulses, the vector potential varies across the sample, requiring the total current densities (TCD) to be calculated by integrating the microscopic current densities (MCD) over the sample. With pulses considered as superposed plane waves, these spatial differences can be translated into variations in the pump-probe delays.
Consequently, TCD at delay $\tau$ is given by
\begin{equation}
J(t,\tau)=\mbox{\large$\int$}\!\mathrm{d}\tau'\,w(\tau'{}\tau)\,j(t,\tau')\,,\label{eq:mismatch}
\end{equation}
with the MCD $ j(t,\tau') $ at delay $ \tau' $, while $ w(\tau'-\tau) $ is a weight function ensuring that only a finite area contributes to the TCD. A detailed derivation is provided in EndMatter part 2.
The pump and probe pulse have a wavelength of $\lambda\,{=}\,1500$\,nm, an electric field amplitude of $F_0\,{=}\,0.004$\,a.u., and a $\cos^2$ envelope with a full width at half maximum of $T\,{=}\,15\,T_0$ (with the pulse period $T_0$). 

\textit{Harmonic-yield oscillation.} The most interesting observables in this setup are the integrated harmonic yields, obtained from TCDs and shown as a function of pump-probe delay $\tau$ in Figs.\,\ref{fig:yield}c and \ref{fig:yield}d. If pump and probe pulses overlap (area I), odd and even harmonics show similar behavior in that both are significantly suppressed. For non-overlapping pulses (area II), the yield oscillates at the optical phonon frequency with an important difference: While odd harmonics oscillate in-phase, which is consistent with the experiment~\cite{refId0}, the even harmonics exhibit out-of-phase oscillations and vary considerably from order to order. The observation suggests that different mechanisms dominate harmonic behavior in area I and II.

\begin{figure}[b]
 \includegraphics[width=8.5cm]{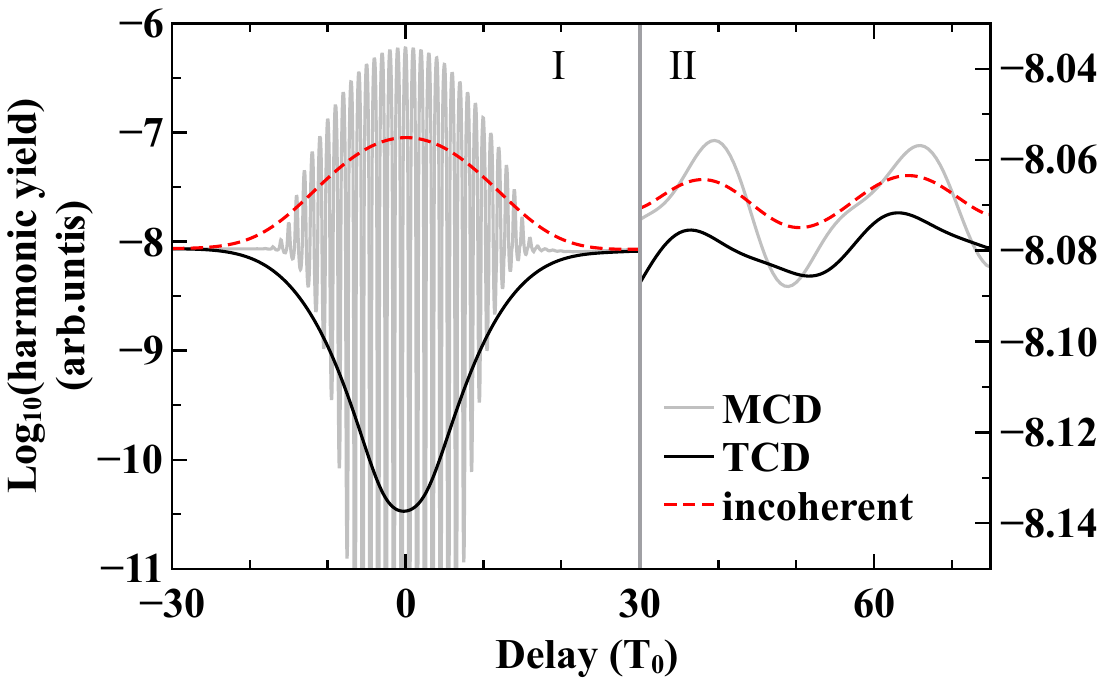}
 \caption{The integrated 12th harmonic yield from MCD (gray curve), its incoherent superposition (red deshed curve) and TCD (black curve), normalized to the integrated yield at asymptotic negative delay ($\tau=-30T_0$) for better comparison. The coordinates of area I is on the left side, while those of area II is on the right side.
}
\label{fig:yield_12th}
\end{figure}%
\textit{Spatial interference effects.} Since the TCDs are derived from the coherent superposition of MCDs at different delays, interference among MCD can significantly impact the harmonic yield from TCD. In area I, the overlap of the two pulses causes swift variations in yield (gray curve in Fig.\,\ref{fig:yield_12th}) and phase (not shown) of the MCD. Hence, interference among the MCD dramatically affects the harmonics, as becomes evident when contrasting the total yield from an incoherent superposition of the MCD versus the true TCD ( black and red dashed curves in Fig.\,\ref{fig:yield_12th}).
In area II of Fig.\,\ref{fig:yield_12th}, the yield fluctuates at the optical phonon frequency. Despite significant differences between MCD and TCD, interference is less critical than in area I. The MCD yield contains a higher frequency component compared with the yield of TCD. This difference arises because higher frequency components are more effectively averaged out by integrating over time delays instead of by interference, as evidenced by similar outcomes from incoherent MCD superposition and TCD.

The suppression of harmonics for temporally overlapping pump and probe pulses (area I) has been observed in many experiments \cite{suppress1,suppress2,suppress3,suppress4,refId0,ex_ph_VO2} and is usually attributed to suppression of interband harmonics caused by the pre-excitation of electrons ~\cite{suppress1,suppress2,suppress3,suppress4}. Our results suggest that spatial interference from the non-coaxial pump-probe setup significantly contributes to harmonic suppression.

\textit{Microscopic origin of yield oscillations.} The strikingly different oscillation features of odd and even harmonics in area II can be traced back to the
fundamental mechanism which generates harmonics and its dependence on symmetries.
The current of an inversion-symmetric system induced by linearly polarized monochromatic light
is antisymmetric under a half-cycle shift $j(t{+}T_0/2)=-j(t)$ ~\cite{Neufeld2019nc,zhangImpactCrystalSymmetries2024a}. Hence, the current can only support odd harmonics. 
In contrast, a system lacking inversion symmetry breaks the half-cycle antisymmetry of $j(t)$ and allows even-harmonic generation. 

\begin{figure}[b]
 \centering
 \includegraphics[width= .75\columnwidth]{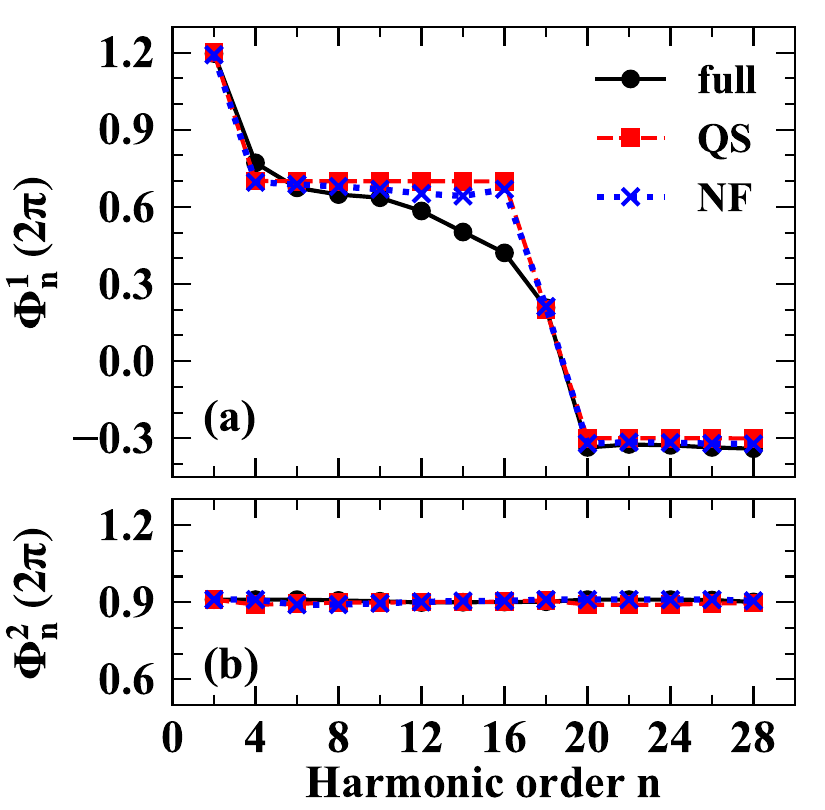}
 \caption{Fitted phases of (a) the fundamental $\Phi_n^1$ and (b) second-harmonic $\Phi_n^2$ components of the phonon frequency in MCD yield oscillations of even harmonics according to Eq.\,\eqref{eq:fitting-model}. 
We show results for full calculations (black circles), as well for the quasi-static (QS, red squares) and the no-feedback (NF, blue crosses) approximations, respectively. Lines are to guide the eye. 
For all harmonics shown, the fit satisfies $R^2 > 0.99$.
}
\label{fig:fitted_phase}
\end{figure}%
In the current pump–probe setup, the half-cycle antisymmetry of the current is statically broken by the lattice distance $D$ and also dynamically due to the lattice motion,
represented by the corresponding velocity $\dot{D}$.
Since the (effective) probe-pulse duration is shorter than the phonon period, we may write the harmonic yield as a function of the lattice dynamics in terms of $\Delta D(\tau)\,{=}\,D(\tau)\,{-}\,D_\mathrm{eq}$ and $\dot{D}(\tau)$
at the peak of the probe pulse, specified by the pump-probe delay $\tau$.
A minimal but faithful description of the yield requires up to 2nd order in $\Delta D$ and 1st order in $\dot{D}$, so that 
\cite{Note1} 
\begin{equation}
\label{eq:even_I}
 I_{n}(\tau) \approx I_{n}^0 + I_{n}^D\,\Delta D(\tau) + I_{n}^{\dot{D}}\,\dot{D}(\tau) + I_{n}^{DD}\,[\Delta D(\tau)]^2
\end{equation}
for even harmonic orders $n{=}2j$.
The displacement $\Delta D(\tau)$ and the velocity $\dot{D}(\tau)$ are out-of-phase by $\pi/2$, forming two orthogonal components analogous to $\cos(\Omega \tau)$ and $\sin(\Omega \tau)$. 
Figure\,\ref{fig:yield}d reveals an order-dependence of the phase or the delay, respectively, of the oscillations in the even harmonic yield which implies that 
 the coefficients $I_{2n}^D$ and $I_{2n}^{\dot{D}}$ depend on the harmonic order.
 To determine this dependence, we parameterize the coefficients regarding their variation with pump-probe delay $\tau$ with a damped Fourier series, consistent with the damping of lattice motion (see EndMatter)
\begin{align}
I_n(\tau) = & C^0_n+C^\sigma_ne^{-2\sigma_n(\tau-\tau_0)}+C_n^1 e^{-\sigma_n(\tau-\tau_0)}\cos(\Omega\tau+\Phi_n^1)\notag\nonumber\\
&+C_n^2 e^{-2\sigma_n(\tau-\tau_0)}\cos(2\Omega\tau+\Phi_n^2),
\label{eq:fitting-model}
\end{align}
where $C_n^m$ and $\Phi^m_n$ denote the amplitudes and phases belonging to the frequencies $m\Omega$, 
$C^\sigma_n$ is the coefficient of the purely decaying mode and $\sigma_n$ is the damping coefficient. Note that damping starts at $\tau_0 = 30T_0$, when the pump and probe pulses separate. 

Indeed, the fitted phase $\Phi^1_n$ of the oscillation with \(\Omega\) to which $\Delta D$ and $\dot D$ contribute depends on the harmonic order $n$, see Fig.\,\ref{fig:fitted_phase}a. In contrast $\Phi^2_n$, to which only $\Delta D^2$ contributes, does not depend on $n$, see Fig.\,\ref{fig:fitted_phase}b.
We note in passing that for odd harmonics, the part of $j(t{+}T_0/2)$ that matches $-j(t)$ is unaffected by dynamic symmetry breaking, leaving their intensities dependent only on $\Delta D$. Consequently, odd-harmonic yields exhibit oscillations that are independent of harmonic order $n$, as apparent in Fig.\,\ref{fig:yield}c. 

The extracted phases of the even harmonics (black points in Fig.\,\ref{fig:fitted_phase}a) depend smoothly on the order $n$ in the ``responsive range'' between orders $n=4$ and $n=18$, bracketed by two phase jumps by $\pi/2$ between orders $2,4$ and $18,20$, respectively. We expect that in the responsive range, the yield is  most sensitive to different approximations of the intricate interplay between the probe pulse, the electron and the phonon dynamics. Indeed, the phases $\Phi^1_n$
resulting from two different approximations to the dynamics, which are commonly used, behave quite differently. 
In the quasi-static approximation (QS) the nuclei are kept fixed during the probe pulse such that dynamical symmetry breaking is absent.
The corresponding phase $\Phi^1_n$ (QS, red squares) differs in the responsive range from 
the smoothly decreasing phase for the dynamically symmetry broken full dynamics (black circles).
Reintroducing nuclear motion, similar to the scenarios in Refs.\,\cite{PhysRevA.106.053116,PhysRevB.106.064303}, by having the lattice oscillate as initiated by the pump pulse but neglecting any feedback restores dynamic symmetry breaking. Nevertheless, this ``no-feedback'' (NF) approximation produces a phase (blue crosses in Fig.\,\ref{fig:fitted_phase}a) which is closer to the one from the QS approximation than to the full dynamic
solution. This somewhat surprising result suggests that in the responsive range the even harmonic yield is sensitive to the probe pulse itself, since this feedback is excluded in both approximations.

\begin{figure}[t]
 \includegraphics[width=.75\columnwidth]{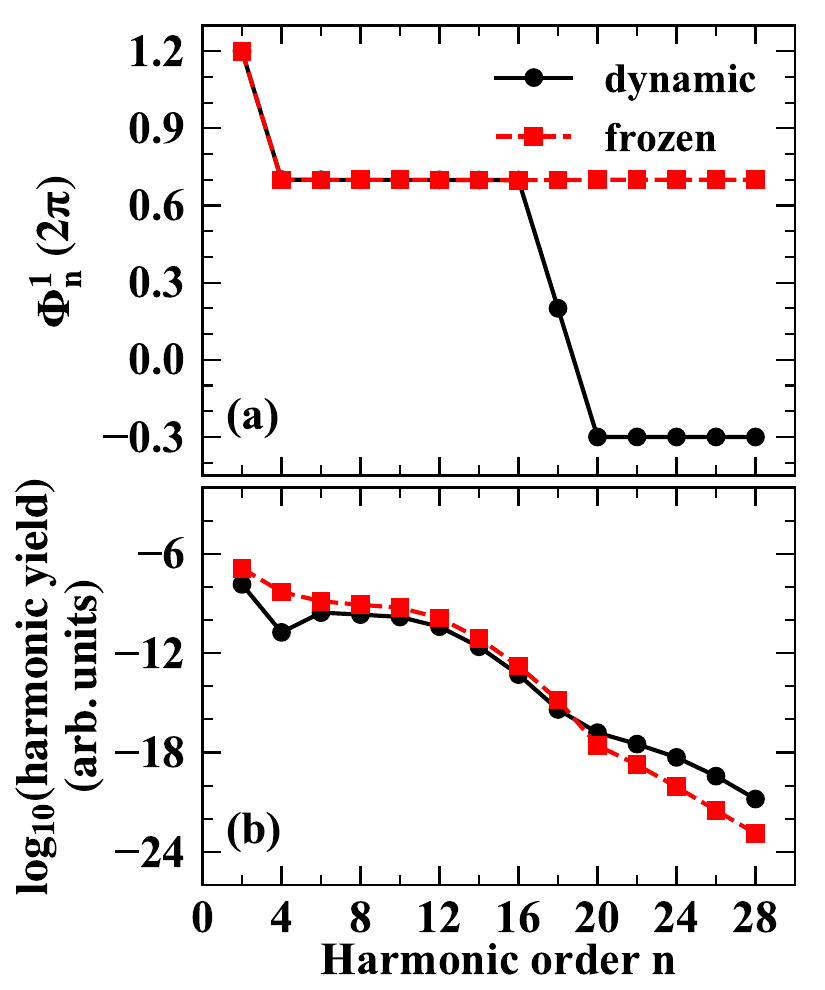}
 \caption{(a) Fitted $\Phi_n^1$ of the fundamental phonon-frequency components in the QS for even harmonics, obtained with dynamic and frozen KS potentials. (b) Integrated even-harmonic yield for the fixed lattice (atoms at equilibrium positions) with dynamic and frozen KS potentials.}
\label{fig:phase_yield}
\end{figure}%
The key difference between fully dynamic and laser driven (NF) nuclei is that atoms tend to move apart due to the probe field in the full simulation.% (see Fig.\,S5 \cite{supp}).
 The $\Phi_n^1$ in the responsive range clearly reflects this subtle difference in lattice dynamics, indicating the great potential of even harmonics to trace subtle features of phonon dynamics, here induced by the probe pulse.

The two phase jumps of $\Phi_n^1$, marking the beginning and the end of the responsive range, are of different origins: 
The phase jump between harmonics $2$ and $4$ marks the transition from intraband to interband harmonic generation. The jump from $18$ to $20$ we interpret as the transition from purely 
laser-driven electron excitation to dominantly electron-electron interaction induced excitation. The results shown in Fig.\,\ref{fig:phase_yield} motivate this interpretation: Freezing the Kohn-Sham (KS) potential at its ground-state shape and thereby excluding electron-electron interaction removes the phase jump (red squares in Fig.\,\ref{fig:phase_yield}a),
which is not the case for a dynamic KS potential (black circles in Fig.\,\ref{fig:phase_yield}a). Moreover,
for harmonics above the 18th order, the yield from the dynamic KS potential is consistently larger, see Fig.\,\ref{fig:phase_yield}b.

\begin{figure}[h!]
 \includegraphics[width=\columnwidth]{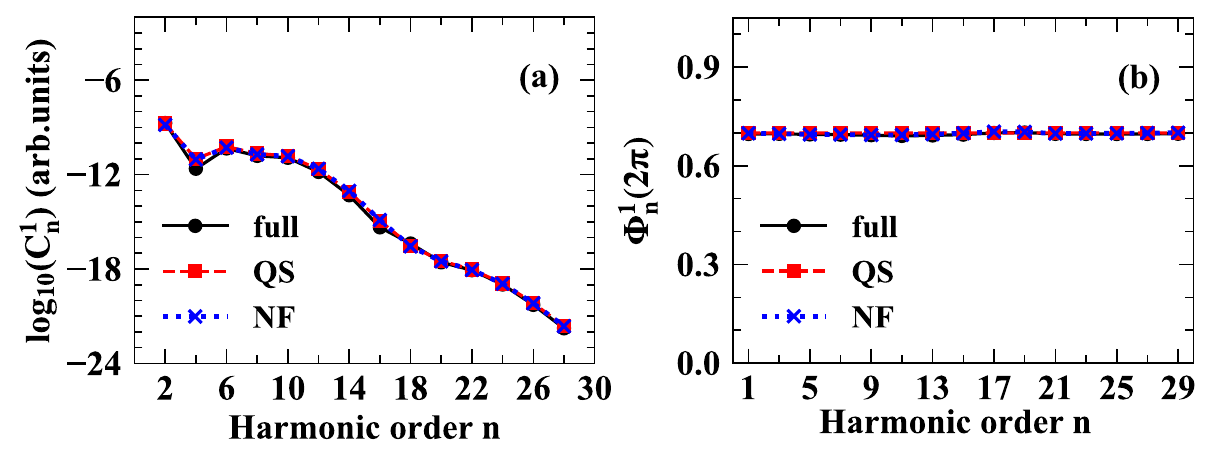}
 \caption{Amplitude $C^1_n$ for even harmonics (a) and phase $\Phi_n^1$ for odd harmonics obtained with Eq.\,\eqref{eq:fitting-model}.}
\label{fig:odd-even-comp}
\end{figure}

We close our analysis by emphasizing that only the phase $\Phi^1_n$ of even harmonics shows the stark sensitivity to details of the dynamics.
All other phases and amplitudes in the parameterization Eq.\,\eqref{eq:fitting-model}, for even as well as for odd harmonics, are not sensitive to the two approximations discussed, as exemplarily shown in Fig.\,\ref{fig:odd-even-comp}.

\emph{Conclusion and Outlook.}
Even harmonics originate from the breaking of half-cycle antisymmetry in the current density and are therefore inherently sensitive to perturbations that induce such symmetry breaking. This sensitivity has been exploited in studies of electron dynamics, such as electron re-collision in gas-phase HHG~\cite{even_g} and the distinction between intraband and interband mechanisms in solids~\cite{Linking_g_s}. Here, we have focused on 
a typical non-coaxial pump-probe setup with the goal of finding the conditions under which even harmonics are most sensitive
to details of electron and coherent phonon dynamics. 
We have confirmed that harmonics are substantially
suppressed if pump and probe pulses overlap, which we attribute mainly to spatial interference induced by the non-coaxial setup, an effect which has not been addressed so far. 

Most importantly, however, by comparing full dynamics of electrons and nuclei
and various commonly used approximations, we have identified a range of even harmonic orders, where the delay of the yield oscillations 
of those harmonics is highly sensitive to subtle dynamical features, namely electron–electron interaction and changes in coherent phonon dynamics, which are in our setup induced by the probe pulse. Our results suggest the experimental detection of even harmonics as a promising tool for uncovering subtle details of phonon and electron dynamics in solid-state systems, when inversion symmetry is statically and dynamically broken. 

\textit{Acknowledgements.}
JL acknowledges a two-year scholarship from the China Scholarship Council to study at the Max-Planck-Institute for the Physics of Complex Systems, where most of the research has been conducted. This work was supported by the National Natural Science Foundation of China (Grant No. 12274188).

\def\articletitle#1{\emph{#1}}
%

%%%%%%%%%%%%%%
% End Matter %
%%%%%%%%%%%%%%
\newpage
\makeatletter 
\renewcommand{\theequation}{E\@arabic\c@equation}
\makeatother
\setcounter{equation}{0}

\section*{End matter}
\subsection{1. Details of the high-harmonics calculations}\noindent
Time-dependent functional theory (TDDFT)  for the electrons is combined with classical nuclear motion, solving the time-dependent Kohn-Sham (KS) equation along with  Hamilton's equations for the nuclei. Due to the translational symmetry of the lattice, the KS orbitals can be expressed as Bloch functions
$|\psi_{n,k}(t)\rangle = e^{\mathrm{i}k x} |U_{n,k}(t)\rangle$,
where $U_{n,k}$ is the wavefunction within a unit cell to which the real-space calculation can be restricted. Substituting $|\psi_{n,k}(t)\rangle$  into the standard time-dependent KS equation gives
\begin{equation}
    \label{eq:KSE}
    \mathrm{i}\partial_t \left| U_{n,k}(t) \right\rangle = (1+\mathrm{i}\gamma) \widehat{H}_{\mathrm{KS}}(k,t) \left| U_{n,k}(t) \right\rangle,
\end{equation}
where $n$ and $k$ denote the orbital index and the crystal momentum, respectively. 
Within the velocity gauge and dipole approximation, the KS Hamiltonian for electrons 
with momentum operator $\widehat{p}_k(t) =-\mathrm{i}\partial_x+k+A(t)$ containing the vector potential $A(t)$
is given by
\begin{align}
\widehat{H}_\mathrm{KS}(k,t) & =\frac{\widehat{p}_k(t)^{2}}{2}-\sum_{\sigma} Q_{\sigma} V(x{-}R_{\sigma})\nonumber\\
& +2\mbox{\large$\int$}\! \mathrm{d}x'\rho(x',t)V(x{-}x')+V_\mathrm{xc}[\rho](x,t),
\end{align}
where $R_{\sigma}$ is the position and $Q_{\sigma}$ is the effective charge  of nucleus $\sigma$ in the cell. 

The electron density $\rho(x,t)=\frac{1}{N_{k}}\sum_{n,k}|U_{n,k}(x,t)|^2$ is sampled with $N_{k}$ points in the first Brillouin zone. 
The sum is taken over the  two bands which are initially occupied.
Given the identical electron densities for spin-up and spin-down states, we only calculate wave functions and electron density for one spin orientation and double the electron density to obtain the total electron density. 
Consequently, a factor of 2 precedes the Hartree potential. The local spin-density (LSD) approximation~\cite{TDDFT_1,TDDFT_2,TDDFT_3} is applied, with the exchange-correlation potential $V_\mathrm{xc}[\rho](x)=\frac{6}{\pi}\rho^{1/3}(x)$.

The Coulombic interaction potential
$V_a(x)=-\frac{1}{\sqrt{x^{2}+a^{2}}}$,
satisfying periodic boundary condition, is
\begin{equation}
\label{Vx}
V(x) = -\frac{4}{L} \sum_{j=1}^{n_V} K_0\left(j\tfrac{2\pi}{L}a\right)\cos\left(j\tfrac{2\pi}{L}x\right),
\end{equation}
with the lattice constant $L=6.4$\,a.u., the smoothing parameter $a=1$\,a.u.\ and $n_V{\to}\infty$. $ K_0 $ is the modified Bessel function of the second kind. For computational efficiency, Eq.\,\eqref{Vx} is evaluated for $n_V\,{=}\,50$ at fixed points, with other values obtained via interpolation. 

The total vector potential $A(t)=\widetilde{A}_{\rm pump}(t{+}\tau)+\widetilde{A}_{\rm probe}(t)$
contains the delay shifted pump and probe pulses of duration $T$, each of them described
by $\widetilde{A}(t)=f(t)A_{0}\sin(\omega_{0} t)$ with peak amplitude $A_0=F_0/\omega_0$
and envelope $f(t) =  \cos^{2}(\pi t/2T)$ for $|t|\leq T$ and zero otherwise.

Classical nuclear coordinates $\{R_\sigma,P_\sigma\}$ follow Hamilton's equations of motion ($\sigma=\mathrm{A,B}$)
\begin{subequations}\label{newton}\begin{align}
\frac{\mathrm{d}R_{\sigma}}{\mathrm{d}t} = {} &\left[P_{\sigma}{-}Q_{\sigma}A(t)\right]/M_{\sigma}, \label{newton1} \\
\frac{\mathrm{d}P_{\sigma}}{\mathrm{d}t} = {} & -\sum_{\sigma'} Q_{\sigma}Q_{\sigma'}V'(R_{\sigma}{-}R_{\sigma'})\nonumber\\
&-\mbox{\large$\int$}\!\mathrm{d}x\, Q_{\sigma}V'(x{-}R_{\sigma})2\rho(x)-\frac{1}{\Gamma_\mathrm{ph}}P_{\sigma},\label{newton2}
\end{align}\end{subequations}
with the nuclear mass $ M_{\sigma} $ and $ V'(x)=\partial_{x}V(x) $. Since phonon damping typically occurs on a picosecond time scale (for instance, $\Gamma_\mathrm{ZnO}\,{=}\,1.6\,$ps  for the transversal optical $E_2^{\text{low}}$ mode in ZnO~\cite{refId0}) a damping of $\Gamma_\mathrm{ph}\,{=}\,0.5\,$ps is used. Varying $\Gamma_\mathrm{ph}$ only affects the damping rate of nuclear motion and does not qualitatively alter other results.

The microscopic current density (MCD) reads
\begin{align}
\label{eq:MCD}  
j(t)= {} & \frac{2}{N_{k_0}}\sum_{n,k}\langle U_{n,k}|\widehat{p}_k(t)|U_{n,k}\rangle
\notag\\ &
+\sum_{\sigma}\frac{Q_{\sigma}[P_{\sigma}-Q_{\sigma}A(t)]}{M_\sigma}.
\end{align}
High-order harmonic spectra are derived from the modulus square of the Fourier-transformed currents. We use 4th-order central differences to discretize the KS equations, sampling 512 points in the first Brillouin zone. 
KS wave functions are evaluated on a grid from [0, 6.4)\,a.u.\ with a grid spacing of 0.05\,a.u., and propagated using the Crank-Nicolson method.

\subsubsection*{Phenomenological dephasing}\noindent
TDDFT does not directly incorporate dephasing terms. This  leads to  harmonic spectra noisier than in the experiment~\cite{PhysRevLett.113.073901, PhysRevB.91.064302,PhysRevA.103.063109} and can obstruct  the analysis of odd and even harmonics. Noise can be reduced with a phenomenological relaxation term $ \mathrm{i}\gamma \widehat{H}_\mathrm{KS}(k_0,t) $ in Eq.\,\eqref{eq:KSE}, where $ \gamma $ controls the relaxation strength. This term propagates wave functions in imaginary time, guiding the system towards its ground state. Due to the non-Hermitian nature of the relaxation term, wave functions do not remain orthonormal during propagation, necessitating the use of the Gram-Schmidt process at each propagation step. Throughout this work we use $\gamma=0.16$. 

\subsubsection*{Ground-state calculation}\noindent
A self-consistent iterative scheme is employed that simultaneously optimizes the electronic structure and the nuclear equilibrium positions. To ensure global charge neutrality, the total number of electrons in the system is fixed at $N_{e} = \sum_\sigma Q_\sigma$, 
being the sum of effective nuclear charges.  The procedure is initialized with a uniform electron density, $\rho(x) = N_e / (2L)$, and a trial set of nuclear coordinates within the unit cell. In each iteration, the KS Hamiltonian is constructed according to the current electron density and nuclear positions, and then diagonalized. The lowest $N_e/2$ energy states are subsequently occupied to generate an updated electron density. With this new density, the nuclear coordinates are evolved according to the damped equation of motion described in Eq.\,\eqref{newton}. This damping process drives the nuclei toward the equilibrium configuration associated with the instantaneous electron density. The cycle repeats until the variations in both the electron density and nuclear positions fall below a specified error tolerance.

To verify the self-consistent calculation, we calculate the total energy of the system for different interatomic distances $D$. The total energy per unit cell is given by 
\begin{equation}
\label{totalE}
\begin{aligned}
E= {} &\frac{2}{N_{k}}\sum_{n,k}\langle U_{n,k}|\frac{[-i\partial_x+k]^{2}}{2}|U_{n,k}\rangle\\
&+\mbox{\large$\int$}\!\mathrm{d}x\,\mathrm{d}x' 2\rho(x)\rho(x')V(x{-}x')+E_\mathrm{xc}[\rho]\\
&+\sum_{\sigma}Q_{\sigma}\mbox{\large$\int$}\!\mathrm{d}x\, 2\rho(x)V(x{-}R_{\sigma})\\
&+\frac{1}{2}\sum_{{\sigma}\neq{\sigma'}} Q_{\sigma}Q_{\sigma'}V\left(R_{\sigma}{-}R_{\sigma'}\right)\,,
\end{aligned}
\end{equation}
where the positions of the two atoms are $R_\mathrm{A,B}\,{=}\,{\pm}D/2$ and the exchange energy is given by $E_\mathrm{xc}[\rho]=-\frac{3}{4}(3/\pi)^{1/3}\int\!\mathrm{d}x\, [2\rho(x)]^{4/3}$~\cite{TDDFT_1}.
As a result, the minimal ground state energy from the self-consistent calculation and  the minimal total energy per unit cell are reached at the same distance $D$
and agree in value $D_\mathrm{eq}$.

\subsection{2. Theoretical treatment of non-coaxial\\ pump-probe measurements}
In non-coaxial pump-probe experiments~\cite{refId0,ex_ph_VO2,zhangHighharmonicSpectroscopyProbes2024}, the propagation directions of the pump and probe beams differ, such that signals in the probe beam direction can be detected cleanly.
However, the microscopic current density (MCD) calculated directly from Eq.\,\eqref{eq:MCD} does not take
the non-coaxial setup into account and includes responses from both,  pump and probe light. Hence, to be consistent with the experiment, it is necessary to remove the system's response to the pump light in the 1D approach we consider.

In order to describe this setup we use the Fourier representation 
$F(t)=\int\!\mathrm{d}\kappa\, F_{\omega_\kappa}\cos(\omega_\kappa t)$
of the pulses  (with pump and probe being the same in our study) whereby $\omega_\kappa=\kappa c$. 
For a thin slab the combined field of both pulses, with the probe pulse being delayed by $\tau$ and inclined by $\theta$ as shown in Fig.\,\ref{fig:yield}, reads
\begin{align}
\tilde F(t,\tau,y)= {} &
\mbox{\large$\int$}\!\mathrm{d}\kappa\, F_{\omega_{\kappa}}\big[\cos(\omega_{\kappa} t)
\notag\\ & \qquad\qquad
+\cos(\omega_{\kappa} [t{+}\tau]-\kappa y\sin\theta)\big].
\end{align}
The dependence on $y$ can be eliminated by rewriting the field as
\begin{subequations}\label{eq:ftauy}\begin{align}
F(t,\tau_{\!y}) & = \mbox{\large$\int$}\!\mathrm{d}\kappa\,F_{\omega_{k}}\big[\cos(\omega_{\kappa} t)+\cos(\omega_{\kappa} [t{+}\tau_{\!y}])\big]
\\ \text{with }
\tau_{\!y}&\equiv\tau-\frac{\kappa y\sin\theta}{\omega_{\kappa}}=\tau-\frac{y\sin\theta}{c}.
\end{align}\end{subequations}
  
Since the solid sample is modeled as an array of parallel, one-dimensional diatomic chains without inter-chain coupling, the microscopic dynamics at a specific position on the sample is determined by the local electric field. The MCD at position \(y\) reads 
\begin{equation}
\tilde j(t,\tau,y)=j(t,\tau_{\!y})
\end{equation}
with $\tau_{\!y}$ given in Eq.\,(\ref{eq:ftauy}b).
The total current density (TCD) of the sample can be expressed as
\begin{equation}
\label{eq:mismatch_phonon}
J(t,\tau)=\mbox{\large$\int$}\!\mathrm{d}\tau_{\!y}\, w(\tau_{\!y}{-}\tau) j(t,\tau_{\!y}).
\end{equation}
The window function $w(\tau)=\exp({-}\tau^{2}/[8T_0]^{2})$
accounts effectively for a finite laser focus.
Modifying the form of $w(\tau)$ does not affect the outcome qualitatively.
Using Eq.\,\eqref{eq:mismatch_phonon}, the spatial integration is effectively recast as an integration over the time-delay variable $\tau_{\!y}$. This eliminates the need to compute the MCD at multiple spatial positions, simplifying the computational process significantly.
\end{document}